\documentclass[prb,aps,superscriptaddress,amsmath,amssymb,showpacs,twocolumn,floatfix]{revtex4-1}
\pdfoutput=1
\usepackage[pdftex]{graphicx}
\usepackage{bm}
\usepackage{bbold}
\usepackage{hyperref}

\newcommand{\bg}{ \begin{gather} }
\newcommand{\eg}{\end{gather}}

\def\K{{\mathcal{K}}}

\newcommand{\Tr}{\mathop{\rm Tr}}

\newcommand{\tr}{\mathop{\rm tr}}

\def\ETh{E_{\text{Th}}}
\def\GN{G}
\def\bb{w}

\begin{document}

\title{Quantum decay of the supercurrent and intrinsic capacitance of Josephson junctions beyond the tunnel limit}

\author{Daniil S. Antonenko}
\affiliation{Moscow Institute of Physics and Technology, Moscow, 141700, Russia}
\affiliation{L. D. Landau Institute for Theoretical Physics, Chernogolovka,
Moscow Region, 142432, Russia}
\author{Mikhail A. Skvortsov}
\affiliation{Skolkovo Institute of Science and Technology, Skolkovo, Moscow Region, 143025, Russia}
\affiliation{L. D. Landau Institute for Theoretical Physics, Chernogolovka,
Moscow Region, 142432, Russia}
\affiliation{Moscow Institute of Physics and Technology, Moscow, 141700, Russia}

\date{\today}

\begin{abstract}
A nondissipative supercurrent state of a Josephson junction is metastable with respect to the formation of a finite-resistance state. This transition is driven by fluctuations, thermal at high temperatures and quantum at low temperatures.
We evaluate the life time of such a state due to quantum fluctuations in the limit when the supercurrent is approaching the critical current.
The decay probability is determined by the instanton action for the superconducting phase difference across the junction.
At low temperatures, dynamics of the phase is massive and is determined by the effective capacitance, which is a sum of the geometric and intrinsic capacitance of the junction.
We model the central part of the Josephson junction either by an arbitrary short mesoscopic conductor described by the set of its transmission coefficients, or by a diffusive wire of an arbitrary length.
The intrinsic capacitance can generally be estimated as $C_*\sim G/E_g$, where $G$ is the normal-state conductance of the junction and $E_g$ is the proximity minigap in its normal part. The obtained capacitance is sufficiently large to qualitatively explain hysteretic behavior of the current-voltage characteristic even in the absence of overheating.

\end{abstract}


\maketitle

\section{Introduction}
Macroscopic quantum tunneling is a fascinating manifestation of quantum-mechanical
behavior in large-scale systems with many degrees of freedom.
It is responsible for a finite life time of a metastable state
that cannot decay classically by thermal activation at zero temperature.
Since 1980-ies, macroscopic quantum tunneling has been studied
in a number of condensed-matter systems:
various types of Josephson junctions,
\cite{CaldeiraLeggett,LO1983-PRB,LO1983-JETP,Kivioja2005,Krasnov2005,Mannik2005,Longobardi2011,dwave-Inomata2005,dwave-Bauch2006,dwave-Longobardi2005,ferro-Massarotti2015}
Josephson junction arrays,\cite{FazioZant}
phase-slip centers,\cite{Giordano1988,Bezryadin2000}
vortices in superconductors,\cite{vortex-review}
small ferromagnetic particles,\cite{magnetic1,magnetic2}
etc.

Theoretical description of macroscopic quantum phenomena
is based on the concept of a collective coordinate, $\chi$.
In the simplest cases, it can be considered as a slow variable,
which allows to integrate out the other electronic degrees of freedom
and end up with an effective quantum mechanics for $\chi(t)$.
The resulting dynamics of the collective degree of freedom
is generically non-local in time since interaction with other
modes produces the retardation effect, intimately related with
dissipation. Intensive studies of dissipative quantum
mechanics\cite{LO1983-JETPL,LO1984-JETP,LO_review}
have been triggered by the pioneering work
of Caldeira and Leggett.\cite{CaldeiraLeggett}

\begin{figure}
\center{
\includegraphics[width=1.0\linewidth]{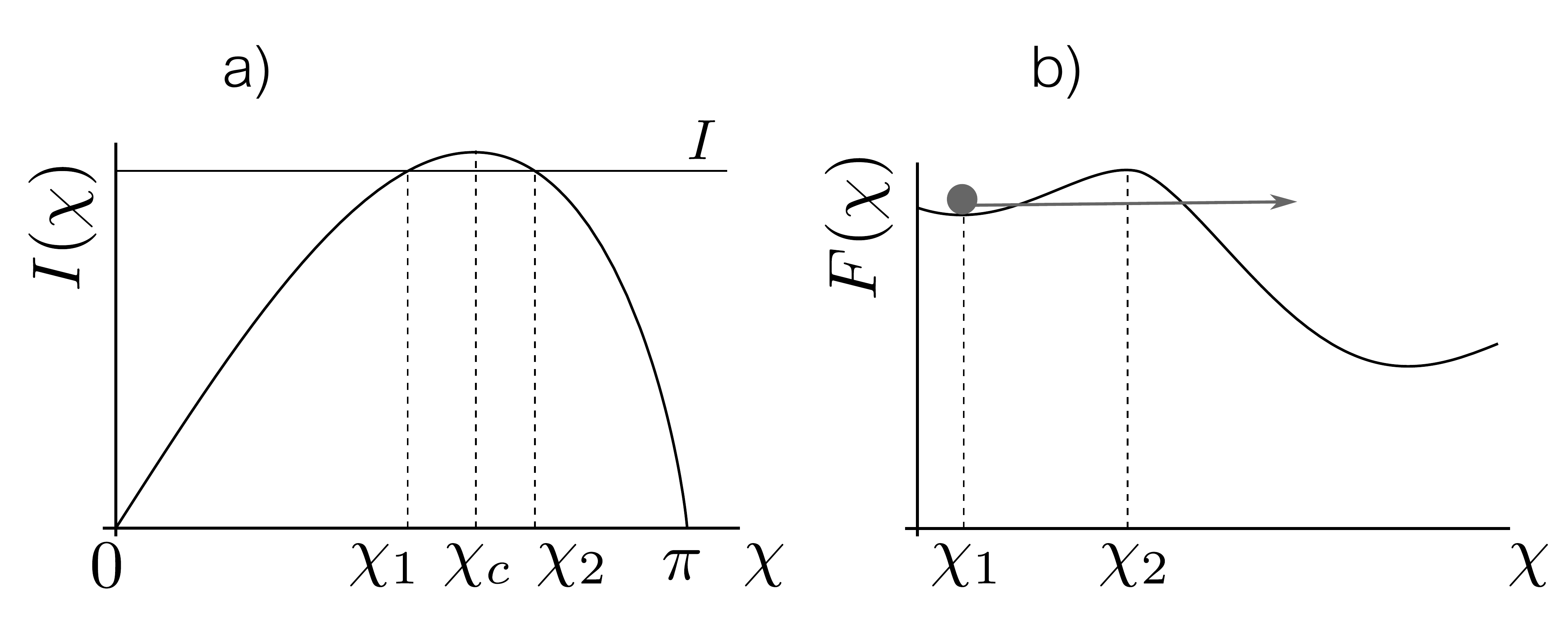}}
\caption{
(a) Typical current-phase relation for a Josephson junction.
(b) Free energy $F(\chi)$ of a current-biased junction vs.\ the phase difference $\chi$, for the current $I$ slightly below $I_c$.}
\label{img:IU_chi_general}
\end{figure}

Among various systems, the Josephson junction can be considered
as a prototypical model of macroscopic quantum tunneling.
Here the superconducting phase difference across the junction, $\chi$,
plays the role of a collective coordinate.
A nondissipative Josephson current can run through the system,
described by a certain current-phase relation $I(\chi)$,\cite{current-phase_relations}
nonsinusoidal in general (Fig.~\ref{img:IU_chi_general}a),
with the critical (maximal) current $I_c$ reached at some
phase difference $\chi_c$. The current can be obtained
by differentiating the free energy of the junction:
$I(\chi) = (2e/\hbar) \, \partial F_0(\chi) / \partial \chi$.
In the current-biased regime with the driving current $I$, the equilibrium states ($\chi=\chi_1\mod2\pi$)
correspond to the minima of the Legendre-transformed free energy,
$F(\chi) = F_0(\chi) - (\hbar/2 e)I \chi$, which has the standard form of a washboard potential
shown in Fig.~\ref{img:IU_chi_general}b.
The supercurrent state with $\chi=\chi_1$
is metastable and does decay (due to quantum or thermal fluctuations) into a resistive branch.
Then the junction usually stays in the dissipative regime unless $I$ is decreased to a smaller retrapping current,
resulting in a hysteretic current-voltage characteristic.\cite{Likharev}

Early studies of the supercurrent decay in Josephson junctions \cite{LO1983-PRB,LO1983-JETP,LO_review} assumed the tunnel limit,
when the superconducting terminals are coupled through an insulating
layer without its own electron dynamics (SIS junctions).
Possible dissipation
and charging
effects were included phenomenologically by adding
an ohmic resistor
and a capacitor
in parallel with the junction [resistively shunted junction (RSJ) model].\cite{Tinkham}
Owing to advances in nanofabrication technology,
current experimental interest has turned towards the study of SNS junctions when two superconducting terminals are connected via a normal region
\cite{Krasnov2005, Dubos2001, Angers08, Crosser08, Krasnov07, Pekola-hysteresis, Meschke2014}
(including graphene\cite{bipolar-graphene,MQT-graphene}, and the surface of a topological insulator\cite{topo_Sacepe, topo_Qu, topo_Veldhorst, topo_Oostinga, topo_Galetti, topo_Kurter} ).
An important new physics in this case is related to the superconducting proximity effect,\cite{PE}
which
renders the normal part of the junction ``partially superconducting''.
The strength of the proximity effect
is characterized by the value of the spectral minigap (at $\chi=0$) which can be
estimated as $E_g\sim\min(\Delta,\hbar/\tau_\text{esc})$, where
$\Delta$ is the superconducting gap in the terminals and $\tau_\text{esc}$
is the time required for an electron in the normal region to establish a contact
with superconductors.\cite{current-phase_relations,Taras-Semchuk-Altland}
For a diffusive wire of length $L$
with good contacts to superconductors, $\hbar/\tau_\text{esc}$ is of the order of the Thouless energy, $\ETh=\hbar D/L^2$,
where $D$ is the diffusion coefficient.

Depending on the relation between $\Delta$ and $\hbar / \tau_\text{esc}$,
one can distinguish between short ($\hbar / \tau_\text{esc} \gg \Delta$, with $E_g\approx\Delta$) and long ($\hbar / \tau_\text{esc} \ll \Delta$, with $E_g\ll\Delta$) Josephson junctions. For short junctions, $I(\chi)$ can be expressed in terms of the transmission coefficients of the normal region,\cite{Beenakker_formula} whereas long junctions require special treatment.\cite{Beenakker-3}
In both cases, the critical current can be written as
\begin{equation}
I_c = i_c \GN E_g/e ,
\label{Ic}
\end{equation}
where $\GN$ is the normal-state conductance of the junction, and $i_c\sim 1$ is a model-dependent factor.
Equation (\ref{Ic}) is a generalization of the Ambegaokar-Baratoff relation to SNS junctions.

\begin{figure}
\center{
\includegraphics[width=0.6\linewidth]{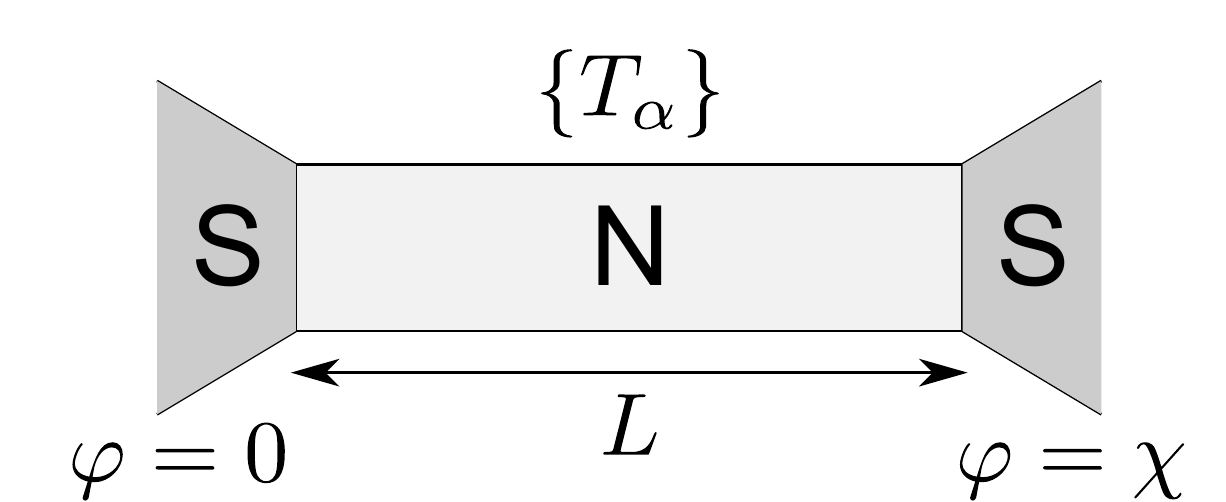}
}
\caption{SNS junction through a normal conductor.
We consider two models of the normal part: (i) an
arbitrary short scatterer with the transmission coefficients $\{T_\alpha\}$ (Sec.~\ref{S:scattering}),
and (ii) a quasi-one-dimensional diffusive wire of an arbitrary length $L$ (Sec.~\ref{S:sigma}).}
\label{img:SNS}
\end{figure}

In this paper we make the first step towards the theory of the supercurrent decay
in Josephson junctions beyond the tunnel limit and calculate the life time
of the supercurrent state due to \emph{quantum}\/ fluctuations.
The normal region of the SNS junction will be modeled either by a short mesoscopic conductor characterized by an arbitrary set of transmission
coefficients $\{T_\alpha\}$ or by a quasi-one-dimensional diffusive wire
of an arbitrary length $L$ (see Fig.~\ref{img:SNS}).
The spectral gaps in superconductors are assumed to be equal.
We will work in the limit $I \rightarrow I_c$
(but not too close to $I_c$ so that the WKB approximation is still applicable).
In this limit free energy $F(\chi)$ is flattened
and $\chi$ becomes the slowest variable in the system, which makes it possible
to treat electronic degrees of freedom in the adiabatic approximation (see the justification in Sec.~\ref{SS:applicability}).

At low temperatures, $T\ll E_g$, thermal quasiparticles responsible for dissipation
are frozen out and the junction can be described
by the imaginary-time effective action with
the capacitive dynamic term:
\begin{equation} \label{our_effective_action}
S[\chi(\tau)] = \int d\tau \left[ \frac{C(\chi_c)}{2 (2 e)^2} \Big( \frac{\partial \chi}{\partial \tau} \Big)^2 + F(\chi) \right] .
\end{equation}
Such a local-in-time description of the phase dynamics is valid only in the supercurrent state with $I\to I_c$.
The effective capacitance,
\begin{equation}
\label{C=C+C}
C(\chi) = C_\text{geom} + C_*(\chi),
\end{equation}
is a sum of the geometric, $C_\text{geom}$, and intrinsic, $C_*(\chi)$, capacitances of the junction.
The latter is determined by the
response of the Andreev bound states to nonstationary boundary conditions.
These states are known to be responsible for carrying the supercurrent.\cite{Nazarov-Blanter}
At lowest temperatures, their low-frequency dynamics cannot be damped (with the kernel $|\omega|$ in terms of Matsubara frequency)
as it is not related to any dissipation, and thus it should be capacitive (with the kernel $\omega^2$).

Thus the problem of quantum decay of a nearly critical supercurrent reduces to calculating the intrinsic capacitance $C_*$ of the junction (evaluated at the critical phase $\chi = \chi_c$).
The intrinsic capacitance of the tunnel (SIS) junction
was obtained in the works of Larkin and Ovchinnikov\cite{LO1983-PRB} and Ambegaokar, Eckern and Sch\"on:\cite{AES}%
\begin{equation}
C_*^\text{tun} (\chi) = (3 \pi \GN/ 32\Delta) (1-\cos\chi)
\label{C-tun}
\end{equation}
(here $\GN$ is the tunnel conductance of the barrier).
This result has been recently
extended to the case of arbitrary short junctions by Galaktionov and Zaikin. \cite{Galaktionov_Zaikin}
They derived the general expression for $C_*^\text{short}(\chi)$ in terms of transmission coefficients $T_\alpha$ [see Eq.~(\ref{Galaktionov_Zaikin_formula})]
and found that, quite generally, $C_*^\text{short}(\chi_c)\sim G/\Delta$.
We will generalize that result further
and show that the magnitude of $C_*(\chi)$ for an 
arbitrary SNS junction
is determined by the ratio of the normal-state 
conductance $G$ to
the minigap $E_g$ induced in the normal part
due to the proximity effect [cf.\ Eq.~(\ref{Ic}) for 
$I_c$]:
\begin{equation}
C_*(\chi) = \alpha_C (\chi) \GN/E_g .
\label{C-general}
\end{equation}
The dimensionless capacitance coefficient $\alpha_C(\chi)$ is model-dependent. At the critical phase difference, $\alpha_C(\chi_c)$ is of the order of one, and we will evaluate it
for an arbitrary short scatterer using Galaktionov-Zaikin theory (see Table \ref{tbl:from_multicharge})
and for a quasi-one-dimensional diffusive wire with the help of the nonlinear sigma model
(see Figs.~\ref{img:Capacity-Eg} and \ref{img:Capacity-chi}).

\begin{table}
\caption{
Various properties of some model Josephson junctions ($E_g=\Delta$ marks short junctions), and the resulting values for the capacitance
coefficient $\alpha_C (\chi_c) = C_*(\chi_c)/(\GN / E_g)$.} \label{tbl:from_multicharge}
\begin{ruledtabular}
\begin{tabular}{ccccc}
& Tunnel & Short wire & Chaotic QD & Long wire \\
\hline
$E_g$ & $\Delta$ & $\Delta$ & $\Delta$ & $3.12\,\ETh$ \\
\hline
$\chi_{c}/(\pi/2)$ & 1 & 1.255 & 1.299 & 1.27 \\
$\mathstrut i_c$ & $\pi/2$ & 2.082 & 2.186 & 3.47 \\
$\mathstrut i_c''$ & $\pi/2$ & 1.706 & 1.791 & 2.92 \\
\hline
$\mathstrut \alpha_C (\chi_c) $
& $3 \pi/32$ & $0.589$ & $0.689$ & 0.905 \\
\end{tabular}
\end{ruledtabular}
\end{table}

In the case when the geometric capacitance can be neglected ($C_\text{geom}\ll C_*$), we obtain that the instanton action,
which determines the life time 
$\tau_0\propto\exp(S)$
 of the supercurrent state, depends only on the dimensionless conductance of the normal part,
\begin{equation}
\label{S-gamma}
S
=
\gamma \frac{\GN}{G_Q}
\Big( 1 - \frac{I}{I_c} \Big)^{5/4} ,
\end{equation}
where $G_Q=e^2/\pi\hbar$ is the conductance quantum,
and the coefficient $\gamma\sim1$ is model-dependent.
Its values for several model Josephson junctions are given by
\begin{equation}
\label{gamma-res}
\gamma
=
\begin{cases}
0.874 & \text{(tunnel barrier);} \\
1.65 & \text{(short diffusive wire);} \\
2.59 & \text{(long diffusive wire);} \\
1.83 & \text{(ballistic chaotic quantum dot).} \\
\end{cases}
\end{equation}

The paper is organized as follows.
The zero-tempe\-ra\-tu\-re intrinsic capacitance of a short SNS junction with an arbitrary normal region specified by its transmission coefficients is discussed in Sec.~\ref{S:scattering} on the basis of Galaktionov-Zaikin formula.
In Sec.~\ref{S:sigma}
we present an approach based on the nonlinear sigma model
which allows us to determine the intrinsic capacitance of the Josephson junction through a diffusive quasi-one-dimensional
normal wire of an arbitrary length, tracing its behavior from the short to long-wire limits.
The resulting expression for the life time of a slightly subcritical current
due to quantum fluctuations is presented in Sec.~\ref{S:lifetime},
where we also discuss the limits of applicability of our approach.
In the concluding Sec.~\ref{S:conclusion} we summarize our results and discuss them in the context of the experimentally observed hysteresis in extended Josephson junctions.
Finally, numerous technical details are relegated to several Appendices.

In our systems of units $\hbar = k_B = 1$.

\section{Short mesoscopic conductor}
\label{S:scattering}

\subsection{Scattering matrix approach}

In this Section we consider the case
when the normal part can be described by the set of its transmission
eigenvalues $T_{\alpha}$ with the distribution function
\begin{equation}
{\cal P}(T)
= \Big< \sum_{\alpha} \delta(T-T_\alpha) \Big> .
\end{equation}
The scattering matrix theory proved to be a powerful and intuitively clear
method for studying quantum transport in mesoscopic conductors.\cite{Beenakker_scattering_matrix}
For normal systems, its most renowned predictions are
the celebrated Landauer formula for the dc conductance, \cite{Landauer_formula}
$
\GN = G_Q \sum_{\alpha} T_{\alpha}
$,
and the expression for the
Fano factor in the theory of shot noise,\cite{Blanter_Buttiker_shot_noise}
$
F = \sum_{\alpha}T_{\alpha} (1 - T_{\alpha}) /\sum_{\alpha} T_{\alpha}
$.

Application of the scattering matrix approach to superconducting
hybrid system is a more delicate issue.\cite{Beenakker_formula,Beenakker_scattering_matrix}
It requires the elements of the normal-state scattering matrix to be energy independent on the relevant energy scale set by the minigap $E_g$.\cite{Beenakker-3}
Since the scattering matrix acquires energy dependence at the scale $\hbar/\tau_\text{esc}$,
such a situation is realized for sufficiently short junctions
with $\hbar / \tau_\text{esc} \gg \Delta$
and hence $E_g\approx\Delta$
(a diffusive SNS junction with ideal interfaces belongs to this class if $\ETh\gg\Delta$).
Then the Josephson current can be found with the help of Beenakker formula:\cite{Beenakker_formula}
\begin{equation}\label{Beenakker_formula}
I(\chi) = \frac{e \Delta^2}{2} \sin{\chi} \sum_{\alpha} \frac{T_{\alpha}}{\epsilon_{\alpha}} \tanh \frac{\epsilon_{\alpha}}{2 T},
\end{equation}
where $T$ is the temperature, and $\epsilon_\alpha$ is the energy of the Andreev bound state in the corresponding channel:
\begin{equation}
\label{epsilon-alpha}
\epsilon_{\alpha} = \Delta \sqrt{1 - T_{\alpha} \sin^2 (\chi/2)}.
\end{equation}

Phase dynamics of a short Josephson junction has been recently studied by Galaktionov and Zaikin.\cite{Galaktionov_Zaikin} With the help of the Keldysh technique they derived an effective action for small phase fluctuations near the equilibrium phase $\chi$.
Slow phase dynamics in the zero-temperature limit is governed by the
capacitance (\ref{C=C+C}), where the
intrinsic capacitance is given by
\begin{multline} \label{Galaktionov_Zaikin_formula}
C_{*}(\chi) = \frac{e^2}{4\Delta} \sum_{\alpha} \Bigg\{ \frac{2 - \left( 2 - T_{\alpha} \right) \sin^2 \left( \chi/2 \right)}{T_{\alpha} \sin^4 \left( \chi/2\right)} \\ {} - \left[1 - T_{\alpha} \sin^2 \left( \chi/2 \right) \right]^{-5/2}
\Bigg[ 2T_\alpha \left( T_\alpha - 2 \right) \sin^2 \left( \chi/2 \right)
\\{}
+ 5
+ T_\alpha + \frac{2 - 2 \left( 1 + 2 T_{\alpha} \right) \sin^2 \left( \chi/2 \right)}{T_{\alpha} \sin^4 \left( \chi/2\right)} \Bigg]
\Bigg\}.
\end{multline}
Since the resulting expression is sufficiently involved, we find it instructive to rederive Eq.~(\ref{Galaktionov_Zaikin_formula}) in the Matsubara representation. Though the general line of the derivation is very similar to that of Ref.~\onlinecite{Galaktionov_Zaikin}, the absence of an additional Keldysh matrix structure makes it possible to track the details of the calculation.
This procedure summarized in Appendix \ref{app:derivation_of_K_for_short_junctions} reproduces Eq.~(\ref{Galaktionov_Zaikin_formula}) obtained by Galaktionov and Zaikin.

\subsection{Intrinsic capacitance for model junctions}

Equation (\ref{Galaktionov_Zaikin_formula}) can be used to evaluate numerically the intrinsic capacitance, $C_*(\chi)$, of various short Josephson junctions.\cite{Galaktionov_Zaikin}
We calculate it at the critical phase, $C_*(\chi_c)$, for three types of structures, with the superconducting terminals coupled through the following links:
\begin{itemize}
\item
a tunnel barrier with all $T_\alpha\ll1$,
\item
a short ($\ETh\gg\Delta$) diffusive wire, with ${\cal P}(T)$ given by Dorokhov distribution,\cite{Dorokhov_distribution}
\begin{equation}
\label{Dorokhov}
{\cal P}(T)
=
\frac{\GN}{2G_Q}
\frac{1}{T \sqrt{1 - T}} ,
\end{equation}
\item
a ballistic chaotic quantum dot,\cite{QD-transmission} with
\begin{equation}
{\cal P}(T)
=
\frac{2\GN}{\pi G_Q}
\frac{1}{\sqrt{T(1 - T)}} .
\end{equation}
\end{itemize}

The results along with some characteristics of these Josephson junctions are presented in Table~\ref{tbl:from_multicharge}.
We see that the capacitance coefficient $\alpha_C(\chi_c)$ defined in Eq.~(\ref{C-general}) is generally of the order of unity, depending on the particular distribution of transmission coefficients ${\cal P}(T)$.

\section{Quasi-one-dimensional wire}
\label{S:sigma}

Here we calculate the intrinsic capacitance of the SNS junction made of a normal diffusive wire coupled to superconductors through highly transparent interfaces.
The scattering-matrix approach used in the previous Section cannot be applied to sufficiently long wires ($\ETh\lesssim\Delta$).
In this limit, the energy dispersion of the scattering matrix at relevant energies $E\sim\ETh$ is not negligible, and the action does not have a simple form of Eq.~(\ref{short_junction_action}).\cite{multicharge_action}
To find $C_*(\chi)$ one then has to use a more general method
of the nonlinear sigma model and study spatially inhomogeneous configurations
of the matrix field $Q$ in the wire.

\subsection{Diffusive sigma model and Usadel equation}

Superconducting proximity effect in a diffusive metal can be conveniently described by the
replica sigma model in the imaginary time \cite{Finkel'stein_replica_sigma_model,Oreg}.
Its action for the quasi-one-dimensional metallic wire of Fig.~\ref{img:SNS}
can be written as
\begin{gather}\label{sigma_model_action}
S[Q] = \frac{\GN}{16 G_Q}
\int_{-1/2}^{1/2} dr \Tr \left[ \left( \nabla Q \right)^2 - 4 \varepsilon \tau_3 Q \right],
\end{gather}
where the spacial coordinate $r$ is measured in the units of the wire length $L$,
summation goes over Matsubara energies
$\epsilon = \pi T (2n+1)$,
and $\varepsilon = \epsilon / \ETh$ with the Thouless energy $\ETh = \hbar D / L^2$.
In general, the field $Q$ subject to the constraint $Q^2 = 1$ acts as a matrix in the replica space, but since the phase difference $\chi$ carries no replica index, the structure of the instanton describing quantum tunneling is trivial in the replica space and we can omit the replica index everywhere.
Then $Q = Q_{\epsilon_1, \epsilon_2}(r)$ becomes a matrix in the Nambu-Gor'kov space (Pauli matrices $\tau_i$), as in Appendix~\ref{app:derivation_of_K_for_short_junctions}. 

The stationary saddle point $Q_0(\epsilon_1, \epsilon_2 ; r) = 2 \pi \delta(\epsilon_1-\epsilon_2) Q_0 (\epsilon_1; r)$ for the sigma-model action (\ref{sigma_model_action}) satisfies the Usadel equation\cite{Usadel}
\begin{gather}
- (Q_0 Q_0')' + \varepsilon [\tau_3,Q_0]=0,
\end{gather}
where prime stands for the derivative with respect to $r$.
Assuming perfect NS interfaces, we write the boundary conditions
with the antisymmetric choice of superconducting phases
as $Q_0(\epsilon;\pm1/2)=Q_S^{(\pm\chi/2)}(\epsilon)$,
where $Q_S^{(\varphi)}(\epsilon)$ is defined in Eq.~(\ref{Q_S_general}).

In the standard parametrization
in terms of the spectral angles $\theta_{\epsilon}(r)$
and $\varphi_{\epsilon}(r)$,
\begin{gather}
Q_0 = \left( \tau_1 \cos \varphi - \tau_2 \sin \varphi \right) \sin \theta + \tau_3 \cos \theta,
\end{gather}
the Usadel equation reduces to two coupled equations:
\begin{subequations}
\label{Usadel_equations}
\begin{gather}
\label{Usadel_equation_1}
( \sin^2 \theta \varphi ' ) ' = 0 ,
\\
\label{Usadel_equation_2}
\theta'' - 2 \varepsilon \sin \theta - (\varphi')^2 \sin \theta \cos \theta = 0 ,
\end{gather}
\end{subequations}
with the boundary conditions at $r=\pm1/2$:
\begin{equation}
\theta(\pm 1/2)
= \theta_S (\epsilon)
= \arctan \Delta/\epsilon,
\qquad
\varphi(\pm 1/2) = \pm \chi / 2.
\end{equation}
The stationary supercurrent is given by
\begin{gather} \label{I_from_Usadel}
I(\chi) = \frac{\pi \GN}{e}
T\sum_\epsilon
\sin^2 \theta_\epsilon(r) \, \varphi'_\epsilon(r)
\end{gather}
[its conservation is guaranteed by Eq.~(\ref{Usadel_equation_1})].
Temperature and length dependence of the critical current of diffusive SNS junctions was studied in Ref.~\onlinecite{Dubos2001}.
Our further analysis will be limited to the $T=0$ case.

\subsection{Perturbative expansion near the saddle point}

In the presence of a time-dependent phase difference across the junction, $\chi(\tau)$, the $Q$ matrices in the leads become functions of two time arguments [cf.\ Eqs.~(\ref{QLR})]:
\begin{equation}
\label{Q-chi-bc}
Q_{R(L)}(\tau_1, \tau_2)
=
e^{\pm i\chi(\tau_1)\tau_z/2}
Q_S^{(0)}(\tau_1-\tau_2)
e^{\mp i\chi(\tau_2)\tau_z/2}
.
\end{equation}
In order to get the action as a functional of $\chi(\tau)$ one has to integrate out $Q(r)$ in the wire with the boundary conditions (\ref{Q-chi-bc}). Near the criticality, $I\to I_c$, variations of $x(\tau)=\chi(\tau)-\chi_c$ on the instanton trajectory are small and hence induced (non-diagonal in energy) deviations of $Q$ from the saddle-point $Q_0$ can be treated perturbatively.

Perturbative expansion of the diffusive sigma-model near the supercurrent state in Josephson junctions has been developed in Ref.~\onlinecite{Houzet_Skvortsov}, and here we
generalized it to the case of nonstationary boundary conditions
(an alternative approach would be to work on the level of the dynamic Usadel equations as it was done in Ref.~\onlinecite{Tikhonov2015}). 
In that paper the authors studied the real part of admittance for the case of the long diffusive SNS junction, while the intrinsic capacitance can be extracted from the low-frequency expansion of its imaginary part.

First we present the stationary saddle point $Q_0(r)$ in the form $Q_0 = U^{-1} \tau_1 U$ with the diagonal-in-energy position-dependent matrix $U_\epsilon(r)=e^{-i\tau_{2}[\pi/4-\theta_\epsilon(r)/2]}e^{-i\tau_{3}\varphi_\epsilon(r)/2}$.
Then we parametrize small fluctuations near $Q_0$ as
\begin{equation} \label{W_parametrization}
Q = U^{-1} \tau_{1}( 1 + W + W^{2}/2 + \ldots )U
\end{equation}
in terms of the field $W$ satisfying $\{ \tau_1, W \} = 0$ and
$W^{\dagger} = - W$.
The former constraint implies $W = c \tau_2 + d \tau_3$, and we combine $c$ and $d$ into a vector object
\begin{equation}
\label{w-cd}
w_{\epsilon_1\epsilon_2}(r)
=
V^{-1}_{\epsilon_1\epsilon_2}(r)
\begin{pmatrix}
c_{\epsilon_1\epsilon_2}(r)
\\
d_{\epsilon_1\epsilon_2}(r)
\end{pmatrix} ,
\end{equation}
where the
unitary
matrix $V$ is introduced to simplify the quadratic action [Eq.~(\ref{expression_for_S_ww_bulk}) below].\cite{Houzet_Skvortsov}
Its explicit form can be found in Appendix \ref{app:S_w_and_S_ww}.
The Pauli matrices acting in the $(c, d)$ space will be referred to as $\Sigma_i$. \cite{com-cd}
Physically, the field $w$ describes soft diffusive modes on top of a superconducting state (in the presence of a supercurrent they can no longer be classified as difusons and cooperons).

Expansion of the action in the powers of $w$ reads:
\begin{gather}
\label{expansion_of_action_in_w}
S[Q] = S[Q_0]
+ S^{(w)}[w]
+ S^{(w^2)}[w]
+ \dots ,
\end{gather}
where each term is a combination of the bulk and boundary contributions:
\begin{gather}
S^{(w^n)}[w]
=
S_\text{bulk}^{(w^n)}[w]
+ S_\text{bound}^{(w^n)}[w]
.
\end{gather}
Since $Q_0$ is the saddle point, the linear term $S^{(w)}$ is totally due to the boundary, but it will contribute to the quadratic-in-$x$ action
[see Eq.~(\ref{quadratic_in_x_part_of_action}) below].
The boundary contributions are given by
\begin{gather} \label{expression_for_S_w}
S_\text{bound}^{(w)}=\frac{i \GN}{2G_{Q}}\int\frac{d\epsilon}{2\pi} \, b_\epsilon^T w_{\epsilon, \epsilon} \Big|_{-1/2}^{1/2},
\\
\label{expression_for_S_ww_boundary}
S_\text{bound}^{(w^2)}
=
\frac{\GN}{4 G_{Q}} \int \frac{d\epsilon_{1}}{2\pi} \frac{d\epsilon_{2}}{2\pi} \,
w_{\epsilon_{1},\epsilon_{2}}^\dagger
B_{\epsilon_{1},\epsilon_{2}}
w_{\epsilon_{1},\epsilon_{2}} \Big|_{-1/2}^{1/2}
,
\end{gather}
where
$w_{\epsilon_{1},\epsilon_{2}}^\dagger \equiv (w_{\epsilon_{1},\epsilon_{2}})^\dagger$,
and the operators $b$ and $B$ read
\begin{gather}
b_{\epsilon}(r)
=
V_{\epsilon, \epsilon}^T
\begin{pmatrix}
-\theta_{\epsilon}' \\
\sin\theta_\epsilon\,\varphi_{\epsilon}'
\end{pmatrix} ,
\\ \label{B_definition}
B_{\epsilon_1,\epsilon_2}(r)
=
\nabla
+ i
\frac{ \cos\theta_{\epsilon_1} \varphi_{\epsilon_1}' +
\cos\theta_{\epsilon_2} \varphi_{\epsilon_2}' }{2}
\Sigma_2
.
\end{gather}
The bulk contribution is given by
\begin{equation} \label{expression_for_S_ww_bulk}
S_\text{bulk}^{(w^2)} = \frac{\GN}{4 G_{Q}} \int \frac{d\epsilon_{1}}{2\pi} \frac{d\epsilon_{2}}{2\pi}
\!
\int_{-1/2}^{1/2} \!dr\,
\bb_{\epsilon_{1}\epsilon_{2}}^\dagger
( - \nabla^2 + \Omega_{\epsilon_1\epsilon_2} )
\bb_{\epsilon_{1}\epsilon_{2}} ,
\end{equation}
where explicit expression for the effective potential $\Omega$ can be found in Appendix \ref{app:S_w_and_S_ww}.

\subsection{Effective capacitance evaluation}

The effective action for the phase variable $x(\tau)$ is determined by the saddle-point trajectory of the action~(\ref{expansion_of_action_in_w})
with the boundary conditions (\ref{Q-chi-bc}).
It can be written as a series in $x$:
\begin{equation}
w=w^{(1)}+w^{(2)}+\dots,
\end{equation}
with $w^{(n)}\propto x^n$.
As we are interested in quadratic-in-$x$ terms, it suffices to follow only $w^{(1)}$ and $w^{(2)}$.
Due to nonlinearity of the theory, the linear terms $w^{(1)}$ may influence the equation of motion for the quadratic term $w^{(2)}$.
However, as shown in Appendix \ref{app:higher_orders}, this mechanism gives no contribution to the effective action. Therefore we may consider linear equations of motion obtained by varying $S_\text{bulk}^{(w^2)}$:
\begin{gather} \label{equations_of_motion_for_b}
\left[ - \nabla^2 + \Omega_{\epsilon_1, \epsilon_2}(r) \right] \bb_{\epsilon_1,\epsilon_2}(r) = 0 ,
\end{gather}
and process each order in $x$ independently.

The boundary conditions for Eq.~(\ref{equations_of_motion_for_b}) can be obtained by comparing Eq.~(\ref{Q-chi-bc}) with the parametrization (\ref{W_parametrization}) at $r=\pm1/2$.
The first two terms are given by
\begin{subequations}
\label{boundary_cond_for_w_in_1_and_2_orders_in_x}
\begin{gather}
\label{boundary_cond_for_w_in_1_order_in_x}
w^{(1)}_{\epsilon_{1}, \epsilon_{2}}(\pm1/2)
=
\xi_{R(L)}^{(1)}(\epsilon_{1}, \epsilon_{2})
x_{\epsilon_{1}-\epsilon_{2}}
,
\\
\label{boundary_cond_for_w_in_2_order_in_x}
w^{(2)}_{\epsilon_{1}, \epsilon_{2}}(\pm1/2)
=
\int\frac{d\epsilon}{2\pi}
\xi_{R(L)}^{(2)}(\epsilon_1, \epsilon_2, \epsilon)
x_{\epsilon_1-\epsilon} x_{\epsilon-\epsilon_2}
,
\end{gather}
\end{subequations}
where
\begin{subequations}
\label{bc-w}
\begin{gather}
\xi_{R(L)}^{(1)}(\epsilon_{1}, \epsilon_{2})
= \mp\frac{i}{2} V_{\epsilon_1, \epsilon_2}^{-1} \left( \pm 1/2 \right)
\begin{pmatrix}
0 \\
\sin\vartheta_{\epsilon_1,\epsilon_2}
\end{pmatrix}
, \\
\xi_{R(L)}^{(2)}(\epsilon_1, \epsilon_2, \epsilon)
= -\frac{i}{16} V_{\epsilon_1, \epsilon_2}^{-1} \left( \pm 1/2 \right)
\begin{pmatrix}
\sin ( \vartheta_{\epsilon_1,\epsilon} + \vartheta_{\epsilon_2,\epsilon} ) \\
0
\end{pmatrix} \! ,
\end{gather}%
\end{subequations}
and
$
\vartheta_{\epsilon,\epsilon'}
=
[\theta_{S}(\epsilon)+\theta_{S}(\epsilon')]/2
$.
In what follows we will need only diagonal elements of $w^{(2)}$, so we introduce the notation $\xi_{R(L)}^{(2)}(\epsilon_1, \epsilon_2) = \xi_{R(L)}^{(2)}(\epsilon_1, \epsilon_1, \epsilon_2)$.

With the boundary conditions (\ref{bc-w}),
the solution of Eq.~(\ref{equations_of_motion_for_b}) can be written in the form:
\begin{subequations}
\label{w_through_xi_r_dep}
\begin{gather}
\label{w_x1_through_xi_r_dep}
w^{(1)}_{\epsilon_{1}, \epsilon_{2}}( r )
=
\xi^{(1)}_{\epsilon_{1}, \epsilon_{2}} (r)
x_{\epsilon_{1}-\epsilon_{2}}
,
\\
\label{w_2_through_xi_r_dep}
w^{(2)}_{\epsilon_{1}, \epsilon_{1}}(r)
=
\int\frac{d\epsilon_2}{2\pi}
\xi^{(2)}_{\epsilon_1, \epsilon_2} (r)
x_{\epsilon_1-\epsilon_2} x_{\epsilon_2-\epsilon_1}
,
\end{gather}
\end{subequations}
where the functions $\xi^{(1)}_{\epsilon_{1}, \epsilon_{2}} (r)$ and $\xi^{(2)}_{\epsilon_1, \epsilon_2} (r)$ obey the same Eq.~(\ref{equations_of_motion_for_b}) as $w_{\epsilon_1, \epsilon_2} (r)$ and satisfy the boundary conditions:%
\begin{subequations}
\label{bc_xi}
\begin{gather}
\label{bc_xi_x1}
\xi^{(1)}_{\epsilon_{1}, \epsilon_{2}} (\pm 1/2) = \xi_{R(L)}^{(1)}(\epsilon_{1}, \epsilon_{2})
,
\\
\label{bc_xi_x2}
\xi^{(2)}_{\epsilon_1, \epsilon_2} (\pm 1/2) = \xi_{R(L)}^{(2)}(\epsilon_1, \epsilon_2) .
\end{gather}
\end{subequations}
To get the quadratic-in-$x$ part of the action, we substitute Eqs.~(\ref{w_through_xi_r_dep}) to Eq.~(\ref{expansion_of_action_in_w}):

\begin{gather} \label{quadratic_in_x_part_of_action}
S^{(2)} = S^{(w^{2})} [ w^{(1)} ] + S^{(w)} [ w^{(2)} ]
,
\end{gather}
which takes the form of Eq.~(\ref{quadratic_terms_in_the_expansion_of_multicharge}) with the kernel ${\cal K}(\epsilon,\omega)$ after the substitution $\epsilon = (\epsilon_1 + \epsilon_2)/2$ and $\omega = \epsilon_1 - \epsilon_2$.
Then the intrinsic capacitance should be determined from Eq.~(\ref{C_through_K}).
To find it numerically, we expand the matrix $\Omega_{\epsilon + \omega / 2, \epsilon - \omega /2}(r)$ and the solutions $\xi^{(n)}_{\epsilon + \omega / 2, \epsilon - \omega /2}(r)$
in powers of $\omega$ as
\begin{gather}
\label{Omega-in-omega}
\Omega_{\epsilon + \omega / 2, \epsilon - \omega /2}(r)
=
\sum_{k=0}^\infty
\Omega_k(\epsilon,r)\omega^k
,
\\
\label{expansion_of_xi_in_omega}
\xi^{(n)}_{\epsilon + \omega / 2, \epsilon - \omega /2}(r)
=
\sum_{k=0}^\infty
\xi^{(n)}_k(\epsilon,r)\omega^k
,
\end{gather}
and
write down the system of linear equations for the first three coefficients:
\begin{align}
& ( - \nabla^2 + \Omega_0 ) \xi^{(n)}_0 = 0, \nonumber \\ \label{three_eqs_on_b}
& ( - \nabla^2 + \Omega_0 ) \xi^{(n)}_1 = - \Omega_1 \xi^{(n)}_0, \\ \nonumber
& ( - \nabla^2 + \Omega_0 ) \xi^{(n)}_2 = - \Omega_1 \xi^{(n)}_1 - \Omega_2 \xi^{(n)}_0,
\end{align}
which should be supplemented by the boundary conditions obtained from the expansion of $\xi_{R(L)}^{(n)}(\epsilon + \omega/2, \epsilon - \omega/2)$ in $\omega$.
Extracting $\partial^2\K(\epsilon,\omega)/\partial\omega^2$ from Eq.~(\ref{quadratic_in_x_part_of_action}), we obtain
\begin{gather}
\frac{\partial^2\K(\epsilon,\omega)}{\partial\omega^2}
\biggr|_{\omega=0}
=
4i b^T \xi_2^{(2)} \Bigr|_{-1/2}^{1/2}
+ 2 \sum_{k=0}^2 \xi_k^{(1)T} \nabla \xi_{2-k}^{(1)} \Big|_{-1/2}^{1/2}
\label{Komegaomega_through_xi}
\end{gather}
[the term originating from the $\Sigma_2$ part of Eq.~(\ref{B_definition}) equals zero].
According to Eq.~(\ref{C_through_K}), the intrinsic capacity of the junction is determined by the integral of Eq.~(\ref{Komegaomega_through_xi}) over $\epsilon$.

\subsection{Notes on numerical evaluation}

The first step of numerical simulation is to find the solutions for the Usadel equations (\ref{Usadel_equations}). One way is to use the explicit expression in terms of elliptical functions, \cite{Houzet_Skvortsov} that reduces to finding two constants from a system of algebraic equations numerically.
However in order to obtain the functions $\theta_{\epsilon}(r)$ and $\varphi_{\epsilon}(r)$ numerically we find it more convenient to follow the other way and to simulate the original differential Eqs.~(\ref{Usadel_equations}) directly.
Obtained spectral angles are used to determine the matrices $\Omega_0$, $\Omega_1$, and $\Omega_2$ in Eq.~(\ref{Omega-in-omega}).
Then the system (\ref{three_eqs_on_b}) is solved numerically, the solutions are substituted to Eq.~(\ref{Komegaomega_through_xi}) and the integral over $\epsilon$
in Eq.~(\ref{C_through_K}) is calculated with a proper energy grid.

As a check of the numerical method, it is instructive to calculate $\int d\epsilon \, \K(\epsilon,0)$. It is easy to see that it should be proportional to the derivative of the current-phase relation $\partial I / \partial \chi$. By checking that this quantity indeed crosses zero at the critical phase $\chi_c$ we can provide an independent test for our numerical calculations.

\begin{figure}
\center{\includegraphics[width=0.98\linewidth]{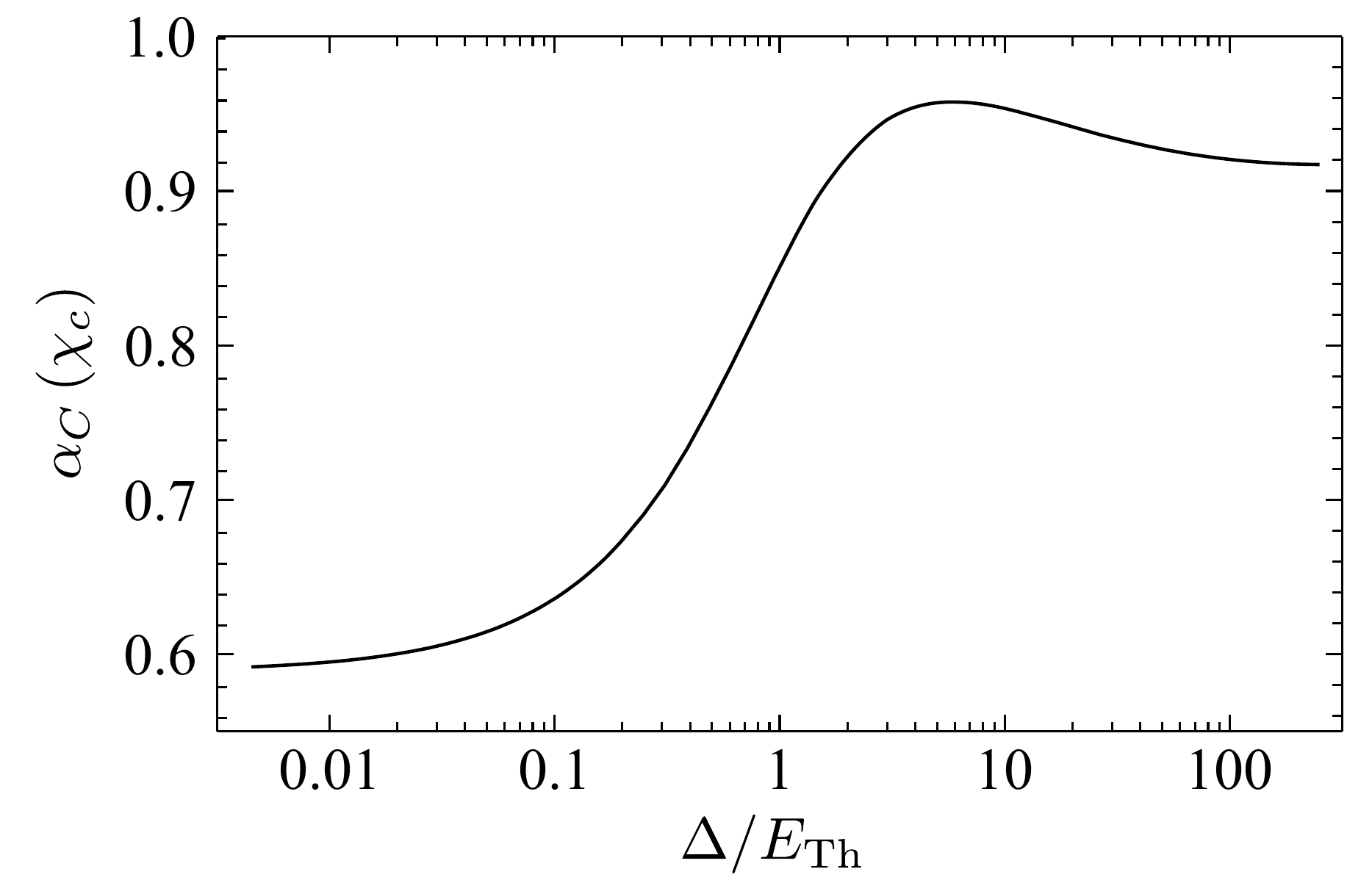}}
\caption{The capacitance coefficient at the critical phase, $\alpha_C(\chi_c)=C_*(\chi_c)/(\GN/E_g)$, of the Josephson junction made of a finite-size wire vs.\ $\Delta / \ETh$.
It is changing from $\alpha_C^\text{short}(\chi_c)=0.589$ for short wires to $\alpha_C^\text{long}(\chi_c)=0.905$ for long wires.
}
\label{img:Capacity-Eg}
\end{figure}

\begin{figure}
\center{\includegraphics[width=0.98\linewidth]{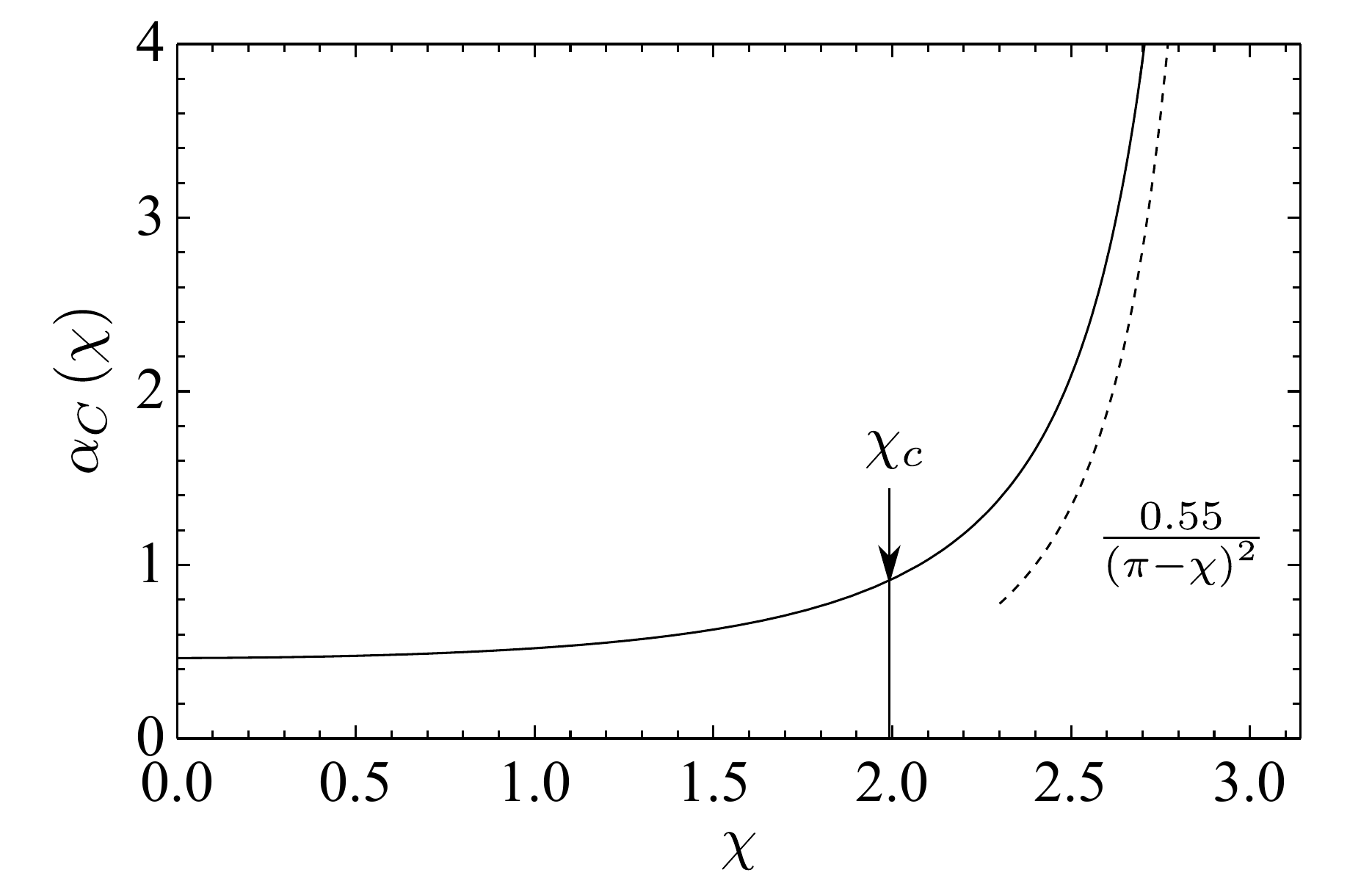}}
\caption{The capacitance coefficient $\alpha_C(\chi)=C_*(\chi)/(\GN/E_g)$ of the long ($\ETh \ll \Delta$) Josephson junction through a quantum wire vs.\ the phase difference across the junction, $\chi$. The critical phase, $\chi_c=1.27 (\pi/2)$, is marked by the arrow.
The dashed line shows the asymptotic behavior (\ref{C-near-pi}) near $\chi=\pi$.}
\label{img:Capacity-chi}
\end{figure}

\subsection{Results}

The resulting dependence of the capacitance coefficient $\alpha_C(\chi_c) = C_*(\chi_c)/(\GN/E_g)$ evaluated at the critical phase difference, $\chi_c$, on the ratio $\Delta/\ETh\propto L^2$ is presented in Fig.~\ref{img:Capacity-Eg} (for the dependence of $E_g$ on $\Delta/\ETh$ see, e.~g., Fig.~5 of Ref.~\onlinecite{Houzet_Skvortsov}).
In general, $\alpha_C(\chi_c)$ is of the order of unity,
varying from $\alpha_C^\text{short}(\chi_c)=0.589$ for short wires
to $\alpha_C^\text{long}(\chi_c)=0.905$ for long wires [in that limit, $E_g=3.12\,\ETh$ (Ref.~\onlinecite{Zhou1998}) and $I_c=10.82\,G\ETh/e$ (Ref.~\onlinecite{Dubos2001})].
The maximal value of
$\alpha_C^\text{max}(\chi_c)=0.958$ is achieved at
$\Delta/\ETh=5.8$, corresponding to the wire length $L=2.4\sqrt{D/\Delta}$.

For completeness, in Fig.~\ref{img:Capacity-chi} we present the phase dependence of the intrinsic capacitance coefficient $\alpha_C(\chi) = C_*(\chi)/(\GN/E_g)$ in the long-wire limit.
In the vicinity of $\chi=\pi$, where the gap closes,
the intrinsic capacitance diverges as
\begin{equation}
\label{C-near-pi}
C^\text{long}_*(\chi) \approx \frac{0.55}{(\pi-\chi)^2} \frac{G_N}{E_g} .
\end{equation}
A similar behavior obtained for short diffusive junctions in Ref.~\onlinecite{Galaktionov_Zaikin} is attributed to the presence of almost open channels ($T_\alpha\to1$) in Dorokhov distribution (\ref{Dorokhov}).

\section{Life time of the supercurrent state}
\label{S:lifetime}

\subsection{Instanton action}

Decay of the dissipationless supercurrent state governed by the action (\ref{our_effective_action}) is equivalent to a quantum-mechanical tunneling of a massive particle under a barrier. Near the criticality, at $I\to I_c$, the potential barrier $\Delta F(x) = F(\chi_c+x) - F(\chi_c)$ can be approximated by a cubic parabola:
\begin{equation}
\label{cubic}
\Delta F(x)
=
\frac{\GN}{G_Q}
\frac{E_g}{2\pi}
\left[
i_c \Big( 1 - \frac{I}{I_c} \Big) x - \frac{i_c''}{6} x^3
\right] ,
\end{equation}
where $i_c$ is defined in Eq.~(\ref{Ic}), and $i_c''=-I''(\chi_c)/(\GN E_g/e)$ is the dimensionless curvature of the current-phase characteristic at $\chi_c$.
Parameters $i_c$ and $i_c''$ for some model Josephson junctions are listed in Table \ref{tbl:from_multicharge}.
The instanton trajectory for the potential (\ref{cubic}) takes the form
\begin{equation}
\chi(\tau) - \chi_1
=
\frac{3 \times (2i_c i_c'')^{1/2} \left( 1 - I/I_c \right)^{1/2} }{\cosh^2 \left( \omega_p \tau / 2 \right)},
\end{equation}
where $\omega_p$ is the plasma frequency describing small phase oscillations near the minimum of $\Delta F(x)$.
In the short-wire limit, $\omega_p$ was calculated in Ref.~\onlinecite{Galaktionov_Zaikin}, and the general expression is given by
\begin{equation}
\label{attempt-freq}
\omega_p = 2^{3/4} (i_c i_c'')^{1/4} \sqrt{ \frac{E_g \GN}{C(\chi_c)}} \Big( 1 - \frac{I}{I_c} \Big)^{1/4}.
\end{equation}

The life time of the subcritical current state in the WKB approximation can be estimated as:
\begin{equation}
\label{tau-result}
\tau_0 \approx (2 \pi/\omega_p) \, e^{S},
\end{equation}
where $S$ is the instanton action,
\begin{equation}
\label{instanton_action}
S = \frac{12 \times 2^{3/4}}{5 \pi} \frac{i_c^{5/4}}{(i_c'')^{3/4}} \frac{\sqrt{C(\chi_c) E_g \GN}}{G_Q}
\Big( 1 - \frac{I}{I_c} \Big)^{5/4},
\end{equation}
This expression generalizes the result for the tunnel limit\cite{LO1983-PRB} to the case of an arbitrary Josephson junction.

Equations (\ref{attempt-freq})--(\ref{instanton_action}) supplemented by the expressions (\ref{C=C+C}) and (\ref{C-general})
provide full description of the supercurrent decay due to quantum fluctuations. The numerical coefficients $i_c$, $i_c''$ and $\alpha_C(\chi_c)$ for a number of model Josephson junctions are summarized in Table \ref{tbl:from_multicharge}.

If the geometric capacitance is sufficiently small, $C_\text{geom}\ll C_*(\chi_c)$, the instanton action (\ref{instanton_action}) can be simplified.
Substituting $C_*(\chi_c)$ from Eq.~(\ref{C-general}), we arrive at Eq.~(\ref{S-gamma}), with the model-dependent parameter
\begin{equation}
\label{gamma-general}
\gamma = \frac{12 \times 2^{3/4}}{5 \pi} \frac{i_c^{5/4} \alpha_C^{1/2}(\chi_c)}{(i_c'')^{3/4}} .
\end{equation}
For some model cases the coefficient $\gamma$ is given by Eq.~(\ref{gamma-res}).

\subsection{Applicability of the theory}
\label{SS:applicability}

In our analysis we rely on the adiabatic approximation for $\chi(\tau)$ justified at $I \rightarrow I_c$. In this limit, variation of the phase is weak and small, that allows to expand the action.
On the other hand, in the very vicinity of $I_c$, the instanton action $S\lesssim1$ and the WKB method fails. Therefore the present theory is applicable as long as
$[ G_Q^2 / C(\chi_c) E_g G ]^{2/5} \ll 1 - I/I_c \ll 1$.
If the geometric capacitance is negligible that reduces to
\begin{equation}
\label{constraint_for_I}
( G_Q / G )^{4/5} \ll 1 - I/I_c \ll 1.
\end{equation}

Finally, we discuss the condition on the temperature range when the $T=0$ description of the phase tunneling is applicable. 
The principal limitation is related to the crossover to thermal decay with the activation exponent $e^{-\delta F/T}$, where $\delta F \sim (G/G_Q) E_g (1 - I/I_c)^{3/2}$ is the height of the free energy barrier.
Quantum description is applicable as long as $\delta F/T\gg S$, which translates into the constraint $[C(\chi_c) T^2 / G E_g]^2 \ll 1 - I/I_c $. In the important limit of a small geometric capacitance, this condition reduces to a simple inequality
\begin{equation} 
\label{constraint_for_T}
  (T/E_g)^4 \ll 1 - I/I_c .
\end{equation}
The condition (\ref{constraint_for_T}) also guarantees that the phase action can still be written in a capacitive form (\ref{our_effective_action}) neglecting quasiparticle damping effects. Indeed, dissipative terms in the action can be roughly described by a large shunting resistance $R(T)\sim G^{-1} e^{E_g/T}$, which has no effect on phase tunneling since the corresponding $RC$ time is much larger than the period of plasma oscillations.

To estimate the predictions of our theory consider a 1\,$\mu$m-long SNS junction similar to that fabricated in Refs.~\onlinecite{Angers08, Pekola-hysteresis, Meschke2014}. With the parameters $G^{-1} \sim 5 \, \Omega$, $\ETh \sim 30$\,mK, and $\Delta/\ETh \sim 10^2$, the intrinsic capacitance is expected to be of the order of $10$\,pF.
The condition (\ref{constraint_for_I}) determines a wide range of allowed bias currents: $10^{-3} \ll 1 - I/I_c \ll 1$, in which the predicted life time 
(when the intrinsic capacitance dominates) 
varies from nanoseconds to practically unlimited values. For example, the decay time $\tau_0 \sim 1$\,s is achieved at $1 - I/I_c \sim 10^{-2}$ . 
According to the constraint (\ref{constraint_for_T}), the quantum tunneling regime is realized in the sub-100\,mK temperature range.

Our calculation was performed in the assumption of ideally transparent SN interfaces, whereas experimentally studied junctions may have a finite conductance $G_T$ of  interfaces. Our theory is valid as long as $G_T \gg G$, but it can be straightforwardly extended beyond this limit by adding the corresponding boundary terms\cite{Efetov-book,Oreg} to the sigma-model action (\ref{sigma_model_action}).
In the opposite limit, $G_T \ll G$, the normal region can be treated as a chaotic quantum dot as it was done in Ref.~\onlinecite{Houzet_Skvortsov}.

\section{Discussion and conclusion}
\label{S:conclusion}

In the present paper we have found the life time of a slightly subcritical dissipationless supercurrent state in an SNS Josephson junction due to quantum fluctuations. At low temperatures, the superconducting phase difference across the junction behaves as a massive quantum-mechanical variable, with the mass determined by the sum of the geometric ($C_\text{geom}$) and intrinsic ($C_*$) capacitances of the junction. While the former describes charging effects, the latter is associated with the dynamics of the Andreev bound states.
We obtain that, generically, the intrinsic capacitance is expressed
through the normal-state conductance $G$ and the spectral minigap $E_g$ induced in the normal part of the junction by the relation $C_*(\chi)=\alpha_C(\chi)(G/E_g)$, where $\alpha_C(\chi)\sim1$ is a model-dependent coefficient.
An analogous expression for $C_*$
in the long-wire limit
was conjectured in
Ref.~\onlinecite{Angers08}, based on the requirement
that the $RC_*$ time of the junction is governed by $\hbar/E_g$.

We determine the model-dependent numerical factor $\alpha_C(\chi_c)$ for two classes of Josephson junctions: an arbitrary short scatterer described by the set of its transmission eigenvalues (see Table \ref{tbl:from_multicharge}) and a diffusive metallic wire of arbitrary length (see Fig. \ref{img:Capacity-Eg}).
In the former case we rely on Galaktionov-Zaikin expression (\ref{Galaktionov_Zaikin_formula}), rederived in Appendix \ref{app:derivation_of_K_for_short_junctions}.
In the latter case we employ the nonlinear sigma model formalism.
Depending on a particular junction, the capacitance coefficient $\alpha_C(\chi_c)$ can vary by a factor of three that may have a pronounced effect on the quantum tunneling rate due to its exponential sensitivity to the junction parameters. For long
junctions with negligible geometric capacitance, the decay rate is determined only by $G$ and proximity to criticality, as given by Eq.~(\ref{S-gamma}).
Measuring the coefficient $\gamma$ in that relation
and comparing it with the results (\ref{gamma-res}) may provide an independent tool to determine the type of the Josephson junction.

Our analysis of the quantum decay of the supercurrent is limited to low temperatures, $T\ll E_g$. For higher temperatures, finite population of quasiparticle states leads to dissipative phase dynamics, and the problem should be treated in the spirit of Ref.~\onlinecite{LO1983-JETP},
with an additional complications due to inapplicability of the simple RSJ model, as well as possible issues on inelastic relaxation and thermalization in the normal part of the junction.\cite{Tikhonov2015}

Capacitive phase dynamics is realized only in the nondissipative supercurrent state near the criticality.
Once it is destroyed by fluctuations, the junction switches to the resistive branch and the phase dynamics becomes dissipative (like for high temperatures).
Though we cannot access this regime, our results can be used for a qualitative description of the hysteretic behavior observed experimentally in lateral junctions.
\cite{Angers08, Crosser08, Krasnov07, Pekola-hysteresis}
Assuming that an oversimpified RSJ model can qualitatively describe the SNS junction with the choice of the resistance $R\sim G^{-1}$ and capacitance $C\sim C_*$, one finds that the McCumber parameter $\beta_C = 2 e I_c C R^2$ 
(which is the square of the quality factor $Q = \omega_p RC$) is generally of the order of one
[formal substitution $R=G^{-1}$ and $C=C_*(\chi_c)$ yields $\beta_C=2i_c\alpha_C(\chi_c)$, which is larger than 6 for long wires].
In this model, large McCumber parameter, $\beta_C>1$, is required for a hysteretic behavior.
Though for long
junctions the geometric capacitance is too small to explain the hysteresis, an account of the intrinsic capacitance provides sufficiently large $\beta_C\gtrsim1$ necessary for observing a hysteretic behavior.\cite{Angers08}
Recently it was demonstrated that low retrapping current is a consequence of electron overheating in the normal region,
when the electron temperature can be several times larger than the bath temperature.\cite{Pekola-hysteresis}
The retrapping current is then identified with the critical current at the elevated electron temperature, explaining large hysteresis. By contrast, we would like to emphasize that the intrinsic junction capacitance
due to dynamics of the Andreev bound states
may itself lead to a hysteretic behavior, even for a perfect thermal contact with the environment.

\acknowledgements
We are grateful to J. P. Pekola, V. V. Ryazanov and K. S. Tikhonov for stimulating discussions.
This work was partially supported by RFBR grant No.\ 13-02-01389.


\appendix

\section{Intrinsic capacitance of a short Josephson junction}
\label{app:derivation_of_K_for_short_junctions}

In this Appendix we rederive Galaktionov-Zaikin formula (\ref{Galaktionov_Zaikin_formula}) in the Matsubara formalism.

\subsection{Matsubara action}
\label{SS:multicharge}

An arbitrary scatterer with an energy-independent scattering matrix
(short-wire limit) sandwiched between two terminals
can be described by the action\cite{Nazarov_circuit, Nazarov_new_circuit,multicharge_action}
\begin{gather} \label{short_junction_action}
S = - \frac{1}{2} \left\langle \sum_{\alpha} \Tr \ln \left[1-\frac{T_\alpha}{4} (Q_L - Q_R)^2 \right] \right\rangle,
\end{gather}
where $Q_L$ and $Q_R$ are the quasiclassical Green functions in the leads.
$Q(\epsilon_1, \epsilon_2)$ is a function of two energy arguments and acts
as a matrix in Nambu-Gor'kov space (N). When multiplying $Q$-matrices and taking trace, integration over $\epsilon_1$ and $\epsilon_2$ is assumed to be done.
The time representation is defined in the conventional way as
\begin{gather}
Q(\tau_1, \tau_2) = \int \frac{d\epsilon_1}{2 \pi} \int \frac{d\epsilon_2}{2 \pi} \, Q(\epsilon_1, \epsilon_2) e^{-i \tau_1 \epsilon_1 + i \tau_2 \epsilon_2} .
\end{gather}
In a stationary uniform superconductor,
$Q_S$ depends only on the time difference,
$Q_S(\tau_1, \tau_2) = Q_S(\tau_1 - \tau_2)$,
and hence is diagonal in the energy representation,
$Q_S(\epsilon_1, \epsilon_2) = 2 \pi \delta (\epsilon_1 - \epsilon_2) Q_S(\epsilon_1)$,
where
\begin{gather}
\label{Q_S_general}
Q_S^{(\varphi)}(\epsilon) =
\frac{1}{\sqrt{\epsilon^2 + \Delta^2}} \begin{pmatrix}
\epsilon & \Delta e^{i \varphi}\\
\Delta e^{- i \varphi} & - \epsilon
\end{pmatrix}_\text{N},
\end{gather}
and $\varphi$ is superconducting phase.

Now we proceed to the derivation of the capacitive term in the action (\ref{our_effective_action}).
We assume zero superconducting phase on the left lead and a time-dependent phase $\chi(\tau)$ on the right lead (see Fig.~\ref{img:SNS}). We assume that $\chi(\tau) = \chi_1 + x(\tau)$, where $x(\tau)$ is a small and slow function of Matsubara time $\tau$:
\begin{subequations}
\label{QLR}
\begin{gather}
Q_L(\tau_1, \tau_2) = Q_S^{(0)}(\tau_1-\tau_2) ,
\\
\tilde{Q}_R(\tau_1, \tau_2) = e^{i \chi(\tau_1) \tau_z/2} Q_L(\tau_1, \tau_2) e^{- i \chi(\tau_2) \tau_z/2} ,
\end{gather}
\end{subequations}
where $\tau_z$ is the Pauli matrix in the N space.
Substituting these expressions into Eq.~(\ref{short_junction_action})
we obtain the action for the phase difference $\chi(\tau)$.

In the limit $I\to I_c$, the free energy barrier protecting the supercurrent state $\chi=\chi_1$ is small, and one can expand the action in powers of $x(\tau) = \chi(\tau)-\chi_c$.
The quadratic term can be written in the form
\begin{equation} \label{quadratic_terms_in_the_expansion_of_multicharge}
S^{(2)} = \frac{\GN}{4 G_Q} \int \frac{d \omega}{2 \pi} \int \frac{d \epsilon}{2 \pi}
\, x_{\omega} x_{-\omega} \, \mathcal{K} (\epsilon, \omega) ,
\end{equation}
with the kernel $\K(\epsilon,\omega)$ calculaed below.
Also, the same limit guarantees that phase dynamics is slow and thus can be described (at $T=0$) by the term $\dot\chi^2$ in Eq.~(\ref{our_effective_action}), with the
intrinsic capacitance of the junction given by
\begin{gather} \label{C_through_K}
C_*
=
\pi \GN \int \frac{d\epsilon}{2 \pi}
\frac{\partial^2\K(\epsilon,\omega)}{\partial\omega^2} \biggr|_{\omega=0}.
\end{gather}

\subsection{Derivation of the kernel $\mathcal{K}$}

Here we calculate the kernel $\mathcal{K}(\epsilon, \omega)$ starting with the action (\ref{short_junction_action}) for an arbitrary distribution of $T_{\alpha}$ (for brevity we omit brackets that denote averaging over $T_{\alpha}$).
We write $\tilde Q_R$ as
\begin{gather}
\tilde{Q}_{R}(\tau_{1},\tau_{2})=e^{ix(\tau_{1})\tau_{z}/2}Q_{R}(\tau_{1},\tau_{2}) e^{-ix(\tau_{2})\tau_{z}/2},
\end{gather}
where $Q_R(\tau_1, \tau_2) = Q_S^{(\chi_1)}(\tau_1-\tau_2)$ corresponds to the stationary phase $\chi_1$.
Then we insert expressions (\ref{QLR}) into Eq.~(\ref{short_junction_action}) and expand the action in powers of $x(\tau) = \chi(\tau)-\chi_1$.

\begin{widetext}
The first variation
of the action
with respect to $x$ reads:
\begin{equation}
\delta^1 S=- \sum_{\alpha} \frac{iT_{\alpha}}{8}\Tr\frac{\Delta\tilde{Q}}{1-(T_{\alpha}/4)\Delta\tilde{Q}^{2}}
[\tau_{z}\hat{x},\tilde{Q}_{R}]
=
- \frac{i}{2} \sum_{\alpha} \sum_{k=0}^{\infty} \left(\frac{T_{\alpha}}{4}\right)^{k+1} \Tr \Delta\tilde{Q}^{2k+1} [\tau_{z}\hat{x},\tilde{Q}_{R}],
\end{equation}
where we denote $\Delta\tilde{Q}=Q_{L}-\tilde{Q}_{R}$ and $\left(\hat{x}\right)_{tt'}=x(t)\delta_{tt'}$.

In the following expression for the second variation we will omit for simplicity terms that contain two $\hat{x}$ operators without any $Q$-matrix in between, since such terms will give no contribution
to the intrinsic capacitance. Also after taking the variation, we put $x(\tau) = 0$
as we are interested in the second variation near the stationary solution. The result can be written as
\begin{multline}
\delta^{2}S = - \sum_{\alpha} \frac{T_{\alpha}}{32}\sum_{k=0}^{\infty}\left(\frac{T_{\alpha}}{4}\right)^{k} \Tr
\Bigg\{
\sum_{l=0}^{2k-1}\Delta Q^{l} [\tau_{z}\hat{x},Q_{R}] \Delta Q^{2k-l}\tau_{z}\hat{x}Q_{R}+\Delta Q^{2k}\tau_{z}\hat{x}Q_{R}\tau_{z}\hat{x}Q_{R}+\Delta Q^{2k+1}\tau_{z}\hat{x}Q_{R}\tau_{z}\hat{x}
\\
-\sum_{l=1}^{2k}\Delta Q^{l} [\tau_{z}\hat{x},Q_{R}] \Delta Q^{2k-l}Q_{R}\tau_{z}\hat{x}+\Delta Q^{2k+1}\tau_{z}\hat{x}Q_{R}\tau_{z}\hat{x}+\Delta Q^{2k}Q_{R}\tau_{z}\hat{x}Q_{R}\tau_{z}\hat{x}
\Bigg\} ,
\end{multline}
where $\Delta Q = Q_L-Q_R$ is diagonal in the energy representation:
\begin{gather}
\Delta Q(\epsilon)=\frac{2\Delta\sin(\chi/2)}{\sqrt{\epsilon^{2}+\Delta^{2}}} q,
\qquad
q = \tau_x \sin(\chi/2) + \tau_y \cos(\chi/2) .
\end{gather}
Switching to the energy representation, we arrive at the action (\ref{quadratic_terms_in_the_expansion_of_multicharge})
with the kernel
\begin{multline}
\mathcal{K}(\epsilon,\omega) = - \sum_{\alpha} \frac{T_{\alpha}G_{Q}}{8G}
\sum_{k=0}^{\infty}\left(\frac{T_{\alpha}}{4}\right)^{k}
\Bigg\{ \sum_{l=0,2,\dots}^{2k} 2 X_{2k,l}
\left[\tr \tau_{z}Q_{R}(\epsilon)\tau_{z}Q_{R}(\epsilon')- \tr \mathbb{1} \right]
\\
+
\sum_{l=1,3,\dots}^{2k-1} X_{2k,l}
\left[ 2 \tr \tau_{z}Q_{R}(\epsilon)q\tau_{z}Q_{R}(\epsilon')q - \tr\tau_{z}Q_{R}(\epsilon)q Q_{R}(\epsilon)\tau_{z}q - \tr\tau_{z}Q_{R}(\epsilon')q Q_{R}(\epsilon')\tau_{z}q\right]
\\ +
X_{2k,0} + X_{2k,2k}
+
X_{2k+1,0}
\tr q\tau_{z}Q_{R}(\epsilon)\tau_{z}
+
X_{2k+1,2k+1}
\tr q\tau_{z}Q_{R}(\epsilon')\tau_{z}
\Bigg\},
\end{multline}
where $\epsilon'=\epsilon-\omega$, and
\begin{equation}
X_{nl} = \frac{[2\Delta\sin(\chi/2)]^{n}}{\left(\epsilon^{2}+\Delta^{2}\right)^{l/2}\left(\epsilon'^{2}+\Delta^{2}\right)^{(n-l)/2}} .
\end{equation}
Evaluating traces one gets:
\begin{multline}
\mathcal{K}(\epsilon,\omega) = - \sum_{\alpha} \frac{T_{\alpha}G_{Q}}{8G}
\sum_{k=0}^{\infty}\left(\frac{T_{\alpha}}{4}\right)^{k}
\Bigg\{
-\sum_{l=0,2,\dots}^{2k} 4 X_{2k,l}\left(1+\frac{\Delta^{2}-\epsilon\epsilon'}{\sqrt{\Delta^{2}+\epsilon^{2}}\sqrt{\Delta^{2}+\epsilon'^{2}}}\right)
\\
+ \sum_{l=1,3,\dots}^{2k-1} 2 X_{2k,l} \left(2\frac{\epsilon\epsilon'-\Delta^{2}\cos\chi}{\sqrt{\Delta^{2}+\epsilon^{2}}\sqrt{\Delta^{2}
+\epsilon'^{2}}}-\frac{\epsilon^{2}+\Delta^{2}\cos\chi}{\epsilon^{2}+\Delta^{2}}-\frac{\epsilon'^{2}+\Delta^{2}\cos\chi}{\epsilon'^{2}+\Delta^{2}}\right)
\\
+2X_{2k,0} + 2X_{2k,2k} + X_{2k+2,1} + X_{2k+2,2k+1}
\Bigg\}.
\end{multline}
Performing summation over $l$ and $k$ one obtains for the kernel $\mathcal{K}(\epsilon,\omega)$:
\begin{multline}
\mathcal{K}(\epsilon,\omega) = \sum_{\alpha} \frac{T_{\alpha}G_{Q}}{8G}
\frac{\epsilon^{2}+\Delta^{2}}{\epsilon^{2}+\Delta^{2}\left(1-T_{\alpha}\sin^{2}\frac{\chi}{2}\right)}\Bigg[-4\frac{\epsilon'^{2}+\Delta^{2}}{\epsilon^{2}-\epsilon'^{2}}\left(1+\frac{\Delta^{2}-\epsilon\epsilon'}{\sqrt{\Delta^{2}+\epsilon^{2}}\sqrt{\Delta^{2}+\epsilon'^{2}}}\right)-2-\frac{\left(2\Delta\sin\frac{\chi}{2}\right)^{2}}{\sqrt{\Delta^{2}+\epsilon^{2}}\sqrt{\Delta^{2}+\epsilon'^{2}}}\\
+
2\frac{\sqrt{\Delta^{2}+\epsilon^{2}}\sqrt{\Delta^{2}+\epsilon'^{2}}}{\epsilon^{2}-\epsilon'^{2}}\left(2\frac{\epsilon\epsilon'-\Delta^{2}\cos\chi}{\sqrt{\Delta^{2}+\epsilon^{2}}\sqrt{\Delta^{2}+\epsilon'^{2}}}-\frac{\epsilon^{2}+\Delta^{2}\cos\chi}{\epsilon^{2}+\Delta^{2}}-\frac{\epsilon'^{2}+\Delta^{2}\cos\chi}{\epsilon'^{2}+\Delta^{2}}\right)\Bigg]
+
\left\{ \epsilon \leftrightarrow \epsilon' \right\}.
\end{multline}
Finally, expanding this expression to the second order in $\omega$ and integrating over $\epsilon$ we arrive at Eq.~(\ref{Galaktionov_Zaikin_formula}) obtained by Galaktionov and Zaikin.\cite{{Galaktionov_Zaikin}}

\end{widetext}

\section{Explicit form of the matrices $V$ and $\Omega$}
\label{app:S_w_and_S_ww}

The rotation by the matrix $V$ in Eq.~(\ref{w-cd}) is introduced in order to get rid of the first derivative in the bulk action $S^{(w^2)}_\text{bulk}$. Its matrix elements are given by \cite{Houzet_Skvortsov}
\begin{equation}
V_{\epsilon_1,\epsilon_2}(r)
=
e^{i\Sigma_2 [\zeta_{\epsilon_1}(r)+\zeta_{\epsilon_2}(r)]/2} ,
\end{equation}
where $\Sigma_i$ are Pauli matrices in the $(c,d)$ space,
and
\begin{equation}
\zeta_{\epsilon}(r) = -\int_0^r ds \, \cos \theta_{\epsilon}(s) \, \varphi_{\epsilon}(s)' .
\end{equation}
After such a rotation, $S^{(w^2)}_\text{bulk}$ acquires the form (\ref{expression_for_S_ww_bulk}), with the matrix potential $\Omega_{\epsilon_1,\epsilon_2}(r)$ given by
(here and below we omit the spacial coordinate $r$ for brevity)\cite{com-cd}
\begin{multline}
\Omega_{\epsilon_1,\epsilon_2}
=
\alpha_{\epsilon_1,\epsilon_2}
-
\rho_{\epsilon_1,\epsilon_2}
\cos ( \eta_{\epsilon_1} + \eta_{\epsilon_2} ) \Sigma_3
\\ {}
+
\rho_{\epsilon_1,\epsilon_2}
\sin ( \eta_{\epsilon_1} + \eta_{\epsilon_2} ) \Sigma_1 ,
\end{multline}
where
\begin{multline}
\alpha_{\epsilon_1,\epsilon_2}
=
\varepsilon_1 \cos \theta_{\epsilon_1}
- \left[ \theta_{\epsilon_1}'^2 + ( \sin \theta_{\epsilon_1} \varphi_{\epsilon_1}' )^2 \right]/4
\\ {} +
\varepsilon_2 \cos \theta_{\epsilon_2}
- \left[ \theta_{\epsilon_2}'^2 + ( \sin \theta_{\epsilon_2} \varphi_{\epsilon_2}' )^2 \right]/4,
\end{multline}
and
\begin{equation}
\rho_{\epsilon_1,\epsilon_2} = \frac{1}{2}
\sqrt{\theta_{\epsilon_1}'^2 + ( \sin \theta_{\epsilon_1} \varphi_{\epsilon_1}' )^2}
\sqrt{\theta_{\epsilon_2}'^2 + ( \sin \theta_{\epsilon_2} \varphi_{\epsilon_2}' )^2} ,
\end{equation}
and the function $\eta_{\epsilon}(r)$ is given by (as usual, $\varepsilon = \epsilon / \ETh$)
\begin{equation}
\eta_{\epsilon}(r) = - 2 \varepsilon \int_0^r ds \, \frac{ \sin^2\theta_{\epsilon}(s) \,\varphi_{\epsilon}'(s)}{\theta_{\epsilon}'^2(s) + [ \sin \theta_{\epsilon}(s) \varphi_{\epsilon}'(s) ]^2} .
\end{equation}

\section{Irrelevance of nonlinear terms in the action (\ref{expansion_of_action_in_w})}
\label{app:higher_orders}

In this Appendix we show that for finding the capacitive effective action (\ref{our_effective_action}) it is sufficient to use the linear equation (\ref{equations_of_motion_for_b}) [resulting from varying only $S_\text{bulk}^{(w^2)}$] and neglect higher order terms in Eq.~(\ref{expansion_of_action_in_w}).

The only possible extra contribution to the quadratic-in-$x$ part of the action can originate from the influence of $w^{(1)}$ on the equation of motion for $w^{(2)}$ arising from cubic nonlinearity in the action (\ref{expansion_of_action_in_w}).
That would induce a correction $\delta\xi^{(2)}$ to $\xi^{(2)}$ which satisfies
Eq.~(\ref{equations_of_motion_for_b}) with a nonzero right-hand side:
\begin{gather}
( - \nabla^2 + \Omega_{\epsilon_1\epsilon_2}) \delta\xi^{(2)}_{\epsilon_1\epsilon_2;i}
=
\int \frac{d \epsilon}{2 \pi} \,
\hat{h}_{\epsilon_1, \epsilon_2, \epsilon}^{ijk}
\xi^{(1)}_{\epsilon_1, \epsilon;j}
\xi^{(1)}_{\epsilon, \epsilon_2;k} ,
\end{gather}
where the indices $i$, $j$, and $k$ refer to the $(c,d)$ space,
and the operator $\hat{h}$ may contain derivatives with respect to $r$.
The crucial point is that the boundary conditions for $\delta\xi^{(2)}$ are trivial:
\begin{gather} \label{boundary_condition_for_correction_to_b_from_higher_orders}
\delta\xi^{(2)}( \pm 1 / 2 ) = 0 ,
\end{gather}
since they are already satisfied by $\xi^{(2)}$ [see Eq.~(\ref{bc_xi_x2})].
When substituted into $S^{(w)}$, $\delta\xi^{(2)}$ could have produced a correction to $\K$.
But since $S^{(w)}$ [Eq.~(\ref{expression_for_S_w})] is proportional to $w$ at the boundary,
condition (\ref{boundary_condition_for_correction_to_b_from_higher_orders}) guarantees the absence of corrections to $\K$.


\begin{references}
\bibitem{com-cd}
Our notation $w = (c, d)^T$ is different from the notation $w = (d, c)^T$ adopted in Ref.~\onlinecite{Houzet_Skvortsov}.
\end{references}
\end{document}